\documentstyle[epsfig,amsfonts,amsbsy,floats,prl,aps]{revtex}

\begin{document}
\twocolumn
\draft
\wideabs{
\title{Measuring the Quantum State\\ of a Large Angular 
Momentum}
\author{G. Klose, G. Smith, and P. S. Jessen}
\address{Optical Sciences Center, University of Arizona, Tucson, 
Arizona 85721}
\date{\today}
\maketitle
\begin{abstract}
We demonstrate a general method to measure the quantum state of an angular
momentum of arbitrary magnitude.  The $(2F+1)\times(2F+1)$ density matrix is completely determined
from a set of Stern-Gerlach measurements with $(4F+1)$ different orientations of the
quantization axis.  We implement the protocol for laser cooled Cesium atoms
in the $6 S_{1/2} (F=4)$  hyperfine ground state and apply it to a 
variety of test states
prepared by optical pumping and Larmor precession.  A comparison of
input and measured states shows typical reconstruction fidelities 
${\cal F}\gtrsim 0.95$.
\end{abstract}
}   % closing bracket for \wideabs{}
%\narrowtext

% body of paper here
The quantum state of a physical system encodes information which can be 
used to predict the outcome of measurements.  The inverse problem was 
mentioned by Pauli already in 1933 \cite{pauli}: is it possible to uniquely determine 
an unknown quantum state by measuring a sufficiently complete set of 
observables on a number of identically prepared copies of the system?  This 
very basic question has gained new relevance in recent years, following the 
realization that systems whose components and evolution are manifestly 
quantum can perform tasks that are impossible with classical devices, such 
as certain computations, secure communication and teleportation 
\cite{quantcalc}.  As we 
harness quantum coherent dynamics for such purposes, the 
development of techniques to accurately control {\em and measure} quantum states 
becomes a matter of practical as well as fundamental interest.  
Reconstruction of a (generally mixed) quantum state based on a record of 
measurements is a nontrivial problem with no general solution \cite{weigert}, but 
system-specific algorithms have been developed and demonstrated 
experimentally in a limited number of cases.  These include light fields 
\cite{smithey}, molecular vibrations \cite{dunn}, electron orbital motion 
\cite{ashburn}, and center-of-mass 
motion in ion traps \cite{leibfried} and atomic beams \cite{kurtsiefer}.
More recently multi-particle 
quantum states have been measured for entangled spin-1/2 systems in NMR \cite{chuang}  
and for polarization-entangled photon pairs \cite{white}.

In this letter we present a new experimental method to measure the unknown 
quantum state for an angular momentum of arbitrary magnitude.  The protocol 
is implemented for laser cooled Cesium atoms in the $6 S_{1/2}(F=4)$ hyperfine ground state, 
and typically reproduces input test states with a fidelity better than 
0.95.  Our work is motivated in part by our ongoing study of quantum 
transport and quantum coherence in magneto-optical lattices, where 
the atomic spin degrees of freedom couple to the center-of-mass motion 
\cite{tunnel}.  
The correlation between spin and motion in this system offers the 
possibility to use the angular momentum as a `meter' in the sense 
introduced by von Neumann \cite{neumann}, to probe the spinor 
wavepacket dynamics.  A similar application has 
been proposed in cavity QED, where an atomic angular momentum can be used 
to read out the quantum state of light in an optical cavity 
\cite{walser}.  
We expect our technique also to provide a powerful experimental tool
to evaluate the performance and error modes of quantum logic gates 
for neutral atoms \cite{brennen}.

\begin{figure}[tb]
    \epsfxsize=3.375in \epsfbox{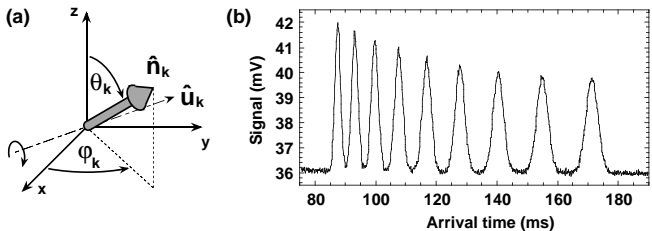}
    \caption{(a) Direction $\hat{n}_{k}$ of a Stern-Gerlach measurement in spherical 
    coordinates. (b) Time-of-flight distribution with well separated 
    peaks corresponding to the populations in the magnetic 
    sublevels $|m\rangle$. }
    \label{F:angle}
\end{figure}

The angular momentum quantum state of an ensemble of atoms with spin quantum 
number $F$ is described by its density matrix $\varrho$.  Newton and Young 
have shown that sufficient information to determine $\varrho$ can be extracted from a 
set of $4F+1$ Stern-Gerlach measurements, carried out 
with different orientations of the quantization axis \cite{newton}. 
They derived an explicit solution for quantization directions 
$\hat{n}_{k}$ with a 
fixed polar angle $\theta$ and evenly distributed azimuthal angles 
$\phi_{k}$.  Our 
reconstruction algorithm uses the same general approach, but takes 
advantage of a simple numerical method to solve for the density matrix for 
a less restrictive choice of directions.  This allows for flexibility in 
the experimental setup and improves the robustness against 
errors.  As a starting point for our reconstruction we choose a space fixed 
coordinate system $\{\hat{x},\hat{y},\hat{z}\}$ in which to determine 
$\varrho$.  If we perform a Stern-Gerlach 
measurement with the quantization axis along $\hat{z}$, we obtain the populations
of the $2F+1$ eigenstates $|m_{z}\rangle$ of $\hat{F}_{z}$, i.e., the diagonal elements of 
the density matrix $\varrho$.  
Information about the off-diagonal elements can be extracted from additional 
Stern-Gerlach measurements with quantization axes along directions 
$\hat{n}_{k}\neq\hat{z}$.  For 
each of these measurements we obtain a new set of populations 
$\pi_{m}^{(k)}$, 
corresponding to the diagonal elements of a matrix $\varrho^{(k)}$ representing 
$\varrho$ in a 
rotated coordinate system with the quantization axis 
oriented along $\hat{n}_{k}$.  The associated coordinate transformation is a rotation 
by an angle $\theta_{k}$ around an axis $\hat{u}_{k}$ in the xy-plane and
perpendicular to $\hat{n}_{k}$ [see Fig.~\ref{F:angle}(a)].  The new 
populations can then be found from a unitary transformation of $\varrho$,
\begin{equation} \label{E:pop}
    \pi_{m}^{(k)} = \langle m| 
   R^{(k) \dagger} \varrho 
    R^{(k)} |m\rangle = \sum_{i,j} R^{(k)*}_{im} R_{jm}^{(k)} \varrho_{ij},
\end{equation}
where $-F\le m,i,j \le F$ and $1\le k \le 4F+1$. The rotation
operators are given by $R^{(k)} = \exp{\left[-\frac{i}{\hbar} \theta_{k} \hat{\bf F} 
\hat{u}_{k}\right] }.$
Arranging all the 
populations $\pi_{m}^{(k)}$ and the density matrix 
elements into vectors $\vec{\pi}$ and $\vec{\rho}$, Eq.~\ref{E:pop}
can compactly be written as $\vec{\pi} = {\mathbf M} \vec{\rho}$, where 
the elements of the rectangular matrix ${\mathbf M}$ are determined by the set of 
rotations $\{ R^{(k)} \}$.  The elements of the density matrix 
$\varrho$ can then be obtained as
$\vec{\rho} = {\mathbf M}^{+} \vec{\pi}$, 
where
\begin{equation} \label{E:pseudo}
     {\mathbf M}^{+} = \sum_{i=1}^{R} \frac{1}{\sqrt{\lambda_{i}}} \:
     {\mathbf w}_{i} \,{\mathbf v}_{i}^{\dagger}
\end{equation}
is the Moore-Penrose pseudoinverse \cite{nrc}. $R$ and 
$\lambda_{i}$ are the rank and non-zero eigenvalues 
of the Hermitian matrix ${\mathbf M}^{\dagger}{\mathbf M}$,
with ${\mathbf w}_{i}$ and ${\mathbf v}_{i}$ being the corresponding eigenvectors of 
${\mathbf M}^{\dagger}{\mathbf M}$ and 
${\mathbf M}{\mathbf M}^{\dagger}$, respectively.

In a real experiment there is both random measurement noise and systematic 
errors, which cause the observed populations $\pi_{m}^{(k)}$ to deviate from those
predicted by Eq.~\ref{E:pop}.  In this situation the pseudoinverse solution
yields a least-squares fit to the data \cite{nrc}.  This fit, however, will become sensitive
to noise and errors if one or more of the $\lambda_{i}$ in 
Eq.~\ref{E:pseudo} are too close to zero. A robust
reconstruction algorithm must therefore use a set of directions 
$\{\hat{n}_{k}\}$ that lead to reasonably 
large $\lambda_{i}$.  A second problem arises because the pseudoinverse solution, while 
always the best fit to the data, is not guaranteed to be a physically 
valid density matrix when noise and errors are present --- a density 
matrix has unit trace, is Hermitian and has non-negative eigenvalues.  The 
pseudoinverse solution automatically fulfills the first two conditions when 
the input populations are normalized. However, when negative eigenvalues occur
we have to employ a different 
method to solve the inversion problem.  For this purpose, we decompose $\varrho$ as
\begin{eqnarray*}
    \varrho = T T^{\dagger} & \quad \text{with} \quad &T_{ij}=0, 
    \text{for } j> i,  \\
      & \quad \text{and} \quad & T_{ii}\in {\mathbb R}, \,\, 
      \sum_{i,j}|T_{ij}|^{2}=1. \nonumber
\end{eqnarray*}
This form automatically enforces the density matrix to have unit trace, 
be Hermitian and positive semi-definite.  
The $(2F+1)^{2}-1$ independent real-valued parameters of the complex, 
lower-triangular matrix $T$ can then be 
optimized to yield the best fit between measured and expected populations.  

We have implemented this general procedure for Cesium atoms in the 
$6 S_{1/2}(F=4)$
hyperfine ground state manifold.  A standard magneto-optical trap (MOT) and optical 
molasses setup is used to prepare an ensemble of $\sim$10$^{6}$ atoms
in a volume of $0.1\text{\, mm}^{3}$ and 
at a temperature of $3.5 \mu\text{K}$.  Atoms from the molasses are loaded into a 
near-resonance optical lattice, composed of a pair of laser beams with 
orthogonal linear polarizations (1D lin$\perp$lin configuration) and 
counter-propagating along the (vertical) z-axis.  Following this second laser
cooling step the atoms are released 
from the lattice, at which point a range of angular momentum quantum states can be 
produced as discussed below.  To allow accurate preparation and 
manipulation of these test states, we measure and compensate the background 
magnetic field in our setup to better than 1~mG.

A laser cooling setup provides a convenient framework for
Stern-Gerlach measurements \cite{enhance}.  At the beginning of each 
measurement we define the quantization axis by applying (switching time 
$\sim$2 $\mu$s) a homogeneous bias magnetic field of $\sim$1~G pointing in the desired 
direction.  All subsequent changes in the magnetic 
field are adiabatic, which ensures that the initial projection of the atomic spin onto 
the direction of the field is preserved at later times.  As the atoms 
fall under the influence of gravity, we apply a strong magnetic field 
gradient ($|{\mathbf B}|\approx 100 \text{ G and } \nabla|{\mathbf 
B}|\approx 100$ G/cm ) by pulsing on the MOT coils for 15 ms.  The resulting 
state dependent force ${\mathbf F}=- m g_{F} \mu_{\text{B}} \nabla
|{\mathbf B}|$ is sufficient to completely separate the arrival 
times for atoms in different $|m\rangle$ as they fall through a probe beam located 
6.9~cm below the MOT 
volume [see Fig.~\ref{F:angle}(b)]. The magnetic populations can then be accurately determined from a 
fit to each separate arrival distribution.  In our case a total of 
$4F+1=17$
Stern-Gerlach measurements are needed to reconstruct $\varrho$.  The
corresponding 17 
different directions $\hat{n}_{k}$ of the bias magnetic field are 
produced by three orthogonal pairs of coils in near-Helmholtz
configuration.  This setup provides complete freedom to choose the 
measurement directions, and allows us to set the dc-magnetic field
with an accuracy of $0.1^{\circ}$ 
in a volume of 1~cm$^{3}.$   For the results reported below
we use 16 measurements on a cone
with $\theta_{k}\approx 82^{\circ}$, and a single measurement with 
$\theta_{k}=0^{\circ}$.

\begin{figure}
    \epsfxsize=3.375in \epsfbox{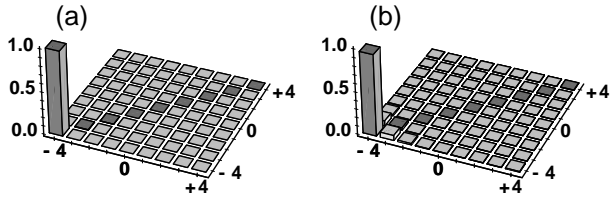}
    \caption{(a) Density matrix of a pure $|m_{z}=-4\rangle$ input state obtained by optical 
    pumping with $\sigma^{-}$-polarized light, and (b) measured  
    density matrix with a fidelity of ${\cal F}=0.97$. Note: All figures display absolute values of the 
    density matrices.}
    \label{F:pm4}
\end{figure}

\begin{figure}[th]
    \epsfxsize=3.375in \epsfbox{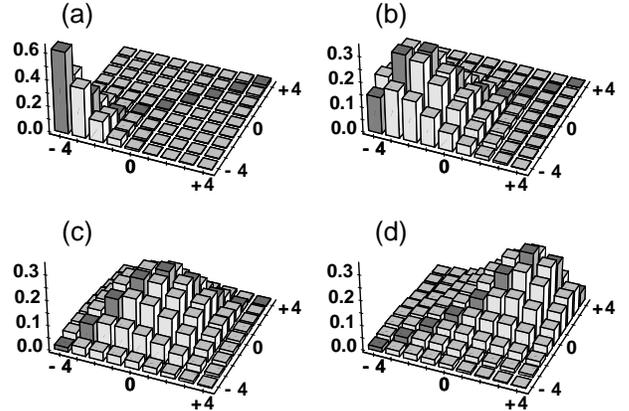}
    \caption{Measured density matrices for spin-coherent states 
    obtained by a controlled Larmor precession of $|m=-4\rangle$ 
    around $\hat{x}$ by (a) 30$^{\circ}$ [${\cal F}=0.95$], (b) 
    60$^{\circ}$ [${\cal F}=0.96$], (c) 
    90$^{\circ}$ [${\cal F}=0.95$], and (d) 120$^{\circ}$ [${\cal 
    F}=0.92$].}
    \label{F:rot}
\end{figure}

We evaluate the performance of our reconstruction procedure by applying it 
to a number of known input states.  These test states are created by a 
combination of laser cooling, optical pumping and Larmor precession in an 
externally applied magnetic field.  First consider optical pumping by a 
$\sigma^{-}$-polarized laser beam tuned to the $6 S_{1/2}(F=4) \to 6 
P_{3/2}(F'=4)$ transition, which allows us to prepare 
the ensemble in a pure $|m_{z}=-4\rangle$ state.  A single Stern-Gerlach
measurement with 
quantization axis along $\hat{z}$ confirms the 
essentially unit population of this state.
For a single non-zero population we can rule out 
off-diagonal elements due to the constraint $|\varrho_{ij}|^{2}\leq 
\varrho_{ii} \varrho_{jj}$, and in this case we therefore know the 
complete density matrix with a high degree of confidence.  Both the input 
and measured density matrices are shown in Fig.~\ref{F:pm4}, and show
excellent agreement.  To quantify the performance we use the 
fidelity \cite{fidelity}
\begin{displaymath}
    {\cal F}=\left(\text{Tr}\left[ \sqrt{\varrho_{i}^{1/2} \varrho_{r} 
    \varrho_{i}^{1/2}}\, \right] \right)^{2},
\end{displaymath}
which is a measure of the closeness between the input ($\varrho_{i}$) and the 
reconstructed ($\varrho_{r}$) density matrices and takes on the value 
${\cal F}=1$ when they are identical.  For an $|m_{z}=-4\rangle$ input 
state our reconstruction fidelity is
${\cal F}=0.97$.  Optical pumping 
with $\sigma^{+}$ yields a nearly pure $|m_{z}=4\rangle$ state, which we can 
similarly reconstruct with a 
fidelity of ${\cal F}=0.94$. Starting from $|m_{z}=-4\rangle$ we can further 
produce a range of spin-coherent 
states \cite{arecchi} by applying a magnetic field along, e.g., the x-axis and 
letting the state precess for a fraction of a Larmor period.  
Figure~\ref{F:rot} shows 
the measured density matrices for four different precession angles, again 
with excellent reconstruction fidelity.

\begin{figure}[ht]
    \epsfxsize=3.375in \epsfbox{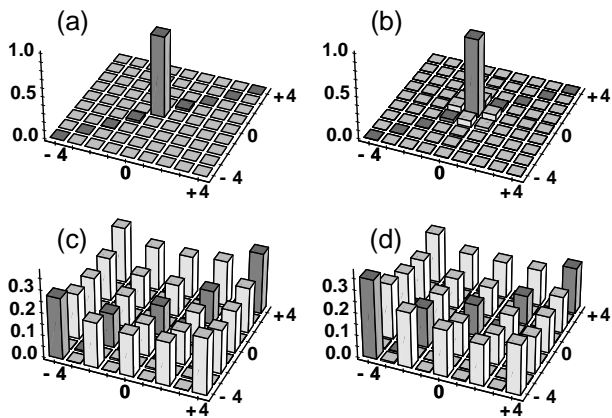}
    \caption{(a) Input state $|m_{y}=0\rangle$ prepared 
    by optical pumping with linear $\hat{y}$-polarized light, and the 
    reconstruction result (b) represented as density matrices in the
    $\{|m_{y}\rangle\}$ basis. The density matrices of input (c) and 
    measured state (d) in the standard $\{|m_{z}\rangle\}$ basis. The 
    fidelity is 0.96.}
    \label{F:mzero}
\end{figure}

\begin{figure}[t]
    \epsfxsize=3.375in \epsfbox{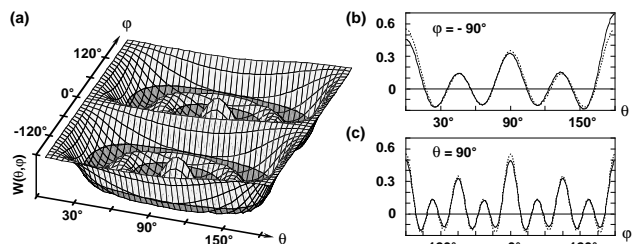}
    \caption{(a) Measured $|m_{y}=0\rangle$ state, represented by the Wigner 
    function $W(\theta,\varphi)$ in spherical phase-space. The darker shading indicates
    negative values. Selected cuts of 
    $W(\theta,\varphi)$ along (b) $\varphi=-90^{\circ}$ and (c) 
    $\theta=90^{\circ}$ (dotted lines = input state, solid lines = 
    measured state).}
    \label{F:wigner}
\end{figure}

It is desirable to also check the reconstruction of test states that are 
not quasi-classical, and whose density matrix exhibits large coherences 
far from the diagonal.  One example of such a state is $|m_{y}=0\rangle$, which we can 
produce by optical pumping on the
$F=4 \to F'=4$ transition with linear polarization along $\hat{y}.$  In 
this basis% with the quantization axis along $\hat{y}$ 
, the 
input test state has the density matrix shown in Fig.~\ref{F:mzero}(a). Its
representation in the %standard
$\{|m_{z}\rangle\}$ basis is easily found by a
(numerical) rotation by $90^{\circ}$ around $\hat{x}$
[Fig.~\ref{F:mzero}(c)], and can be directly 
compared to the reconstruction [Fig.~\ref{F:mzero}(d)].
To simplify the visual comparison we finally rotate the measured density 
matrix by $-90^{\circ}$ around $\hat{x}$ to find its representation
in the $\{|m_{y}\rangle\}$ basis [Fig.~\ref{F:mzero}(b)].
The basis independent reconstruction fidelity 
is  ${\cal F}=0.96$. The non-classical nature of the measured state is 
apparent in its Wigner function representation \cite{dowling}, which takes on 
negative values as shown in Fig.~\ref{F:wigner}. 

\begin{figure}
    \epsfxsize=3.375in \epsfbox{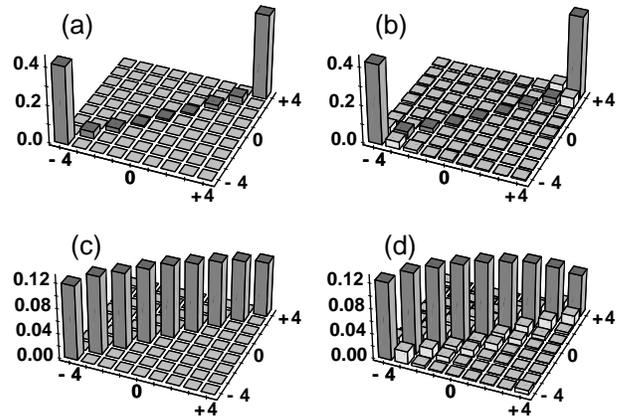}
    \caption{Mixed states: (a) input state from a 
    near-resonance lattice, and (b) measured result. The 
    apparent coherences are real and due to a 6.5$^{\circ}$ tilt of 
    the lattice with respect to $\hat{z}$. (c) Nearly maximally mixed 
    state obtained from optical molasses phase, and (d) corresponding 
    measured density matrix. Matrix comparisons of these mixed states
    each yield a fidelity of 0.99, however, the input state is known 
    with less confidence.}
    \label{F:nrlmol}
\end{figure}

As a final test we have applied our procedure to the (presumably) mixed 
states that result from laser cooling.  Figures~\ref{F:nrlmol}(a,b) show the input and 
reconstructed states produced by laser cooling in a 1D lin$\perp$lin lattice.  
The input state shown here is based on a single Stern-Gerlach 
measurement of the magnetic populations, but it seems reasonable to assume 
that the highly dissipative laser cooling process destroys coherences 
between magnetic sublevels and that the density matrix is
diagonal.  Our measurement confirms this assumption. We can perform the 
same experiment with atoms released directly from a 3D optical molasses, in 
which case there is no preferred spatial direction and one 
expects something close to a maximally mixed state. Figures~\ref{F:nrlmol}(c,d)
show input and measured
density matrices in good agreement with this assumption.

In summary we have demonstrated a method to experimentally determine the 
complete density matrix of a large angular momentum, and tested its 
performance with a variety of angular momentum quantum states of an ensemble 
of laser cooled Cesium atoms in the $6 S_{1/2}(F=4)$ hyperfine ground state.
The input and 
reconstructed density matrices typically agree with a fidelity of ${\cal 
F}\gtrsim 0.95$.  
Limitations on the accuracy to which the density matrix can be measured 
appear to derive from $\approx 3\%$ uncertainties in the individual population 
measurements, and from variations in the direction of the bias magnetic 
field during the first few $\mu$s when it is turned on.  These variations are 
likely due to induced currents in metallic fixtures and additional 
coils in the vicinity of our glass vacuum cell, and translate into 
uncertainty about the exact orientation of the quantization axes for the 
Stern-Gerlach measurements.  Efforts are underway to eliminate these 
problems and achieve even better reconstruction 
fidelities.

We would like to thank K.~Cheong, D.~L.~Haycock, H.~H.~Barrett, 
and the group of I.~H.~Deutsch at the University of New Mexico
for helpful discussions.  This research
was supported by the 
NSF (9732612/9871360), ARO (DAAD19-00-1-0375) and JSOP (DAAD19-00-1-0359).

% now the references. delete or change fake bibitem. delete next three
%   lines and directly read in your .bbl file if you use bibtex.


\end{document}